\documentclass[prb,aps,amsmath,preprint,]{revtex4}
\usepackage{graphicx}

\usepackage{dcolumn}
\usepackage[dvips]{color}

\usepackage{bm}
\usepackage{subfigure}

\begin{document}
\title{High Pressure Study of Lithium Azide from Density-Functional Calculations}

\author{K. Ramesh Babu$^a$, Ch. Bheema Lingam$^b$,
 Surya P. Tewari$^{a,b}$  and G. Vaitheeswaran$^{a,*}$}

\affiliation{$^a $Advanced Centre of Research in High Energy Materials (ACRHEM),
University of Hyderabad, Prof. C. R. Rao Road, Gachibowli, Andhra Pradesh, Hyderabad- 500 046, India
\\$^b$ School of Physics, University of Hyderabad,  Prof. C. R. Rao Road, Gachibowli, Andhra Pradesh, Hyderabad - 500 046, India
}
\date{17 March 2011}
\vspace{0.3in}

\begin{abstract}
\scriptsize {The structural, electronic, optical and vibrational properties of LiN$_3$ under high pressure have been studied using plane wave pseudopotentials within the generalized gradient approximation for the exchange and correlation functional. The calculated lattice parameters agree quite well with experiments. The calculated bulk modulus value is found to be 23.23 GPa which is in good agreement with the experimental value of 20.5 GPa. Our calculations reproduce well the trends in high pressure behavior of the structural parameters. The present results show that the compressibility of  LiN$_3$ crystal is anisotropic and the crystallographic b-axis is more compressible when compared to a- and c-axis which is also consistent with the experiment. Elastic constants are predicted which still awaits experimental confirmation. The computed elastic constants clearly shows that LiN$_3$ is a mechanically stable system and the calculated elastic constants follows the order C$_{33}$ $>$ C$_{11}$ $>$ C$_{22}$ implies that the LiN$_3$ lattice is stiffer along c-axis and relatively weaker along b-axis. 
 Under the application of pressure the magnitude of the electronic band gap value decreases, indicating that the system has the tendency to become semi conductor at high pressures. The optical properties such as refractive index, absorption spectra and photo conductivity along the three crystallographic directions have been calculated at ambient as well as at high pressures. The calculated refractive index shows that the system is optically anisotropic and the anisotropy increases with increase in pressure. The  observed peaks in the absorption and photo conductivity spectra are found to shift towards the higher energy region as pressure increases which imply that in LiN$_3$ decomposition is favored under pressure with the action of light. The vibrational frequencies for the internal and lattice modes of LiN$_3$ at ambient conditions as well as at high pressures are calculated from which we predict that the response of the lattice modes towards pressure is relatively high when compared to the internal modes of the azide ion.}

\end{abstract}
\maketitle
\section{INTRODUCTION}
Alkali metal azides are an interesting class of compounds which find wide range of applications as explosives and photographic materials. These are model systems for studying the fast reactions in solids with complex chemical bonding \cite {Fair, Bowden}. Under the action of heat, light these metal azides become unstable and decompose into metal and nitrogen. Since the decomposition of metal azides involves the electron-transfer mechanism it would be necessary to understand the electronic band structure of these systems \cite{Evans}. A number of studies at the level of Hartree-Fock (HF) and density functional theory (DFT) has been performed to understand the electronic band structure and decomposition mechanism \cite{Seel, Gordienko, Gord, Younk, Zhu}. Moreover, the high pressure behavior of metal azides is an important aspect because of the formation of polymeric nitrogen, a high energy density material \cite{Eremets, Erem}. Previously the high pressure study on sodium azide, NaN$_3$, revealed that a new structure is formed with nitrogen atoms connected by single covalent bonds, which was considered to be an amorphous like structure \cite{Ere}. Since NaN$_3$ and LiN$_3$ are isostructural at ambient conditions, it would be of interest to study the pressure effect on LiN$_3$ with the motive of formation of polymeric nitrogen. Recently {Medvedev et al \cite{Medvedev} reported the behavior of LiN$_3$ under high pressures. Their study revealed that the system is stable up to the pressure of 60 GPa, which is in contrast to that of sodium azide that undergoes a set of phase transitions below the pressures of 50 GPa \cite{Medvedev}. With this motivation we aim to study theoretically the pressure effect on LiN$_3$ crystal system. In order to understand the behavior of LiN$_3$ under high pressure, it is essential to study the physical and chemical properties of the system under high pressure. To the best of our knowledge the structural, electronic, vibrational and optical properties of LiN$_3$ at high pressures have not been explored theoretically. Moreover, these properties under pressure are important in a better understanding of the stability of LiN$_3$.
In this paper, we present a first-principles study of solid monoclinic LiN$_3$ under hydrostatic pressure of up to 60 GPa using density functional theory. The structural parameters, bulk modulus, energy band gap, density of states, optical and vibrational properties of LiN$_3$ under pressure are reported. The remainder of the paper is organized as follows: A brief description of our computational method is given in section 2.
The results and discussion are presented in section 3. Finally we end with a brief summary of our conclusions in section 4.
 \begin{figure}
\centering
\includegraphics[width=10cm]{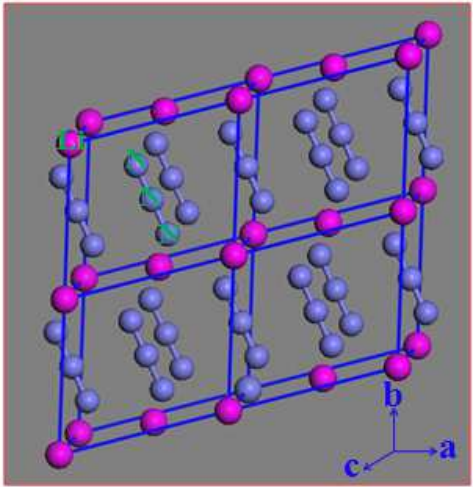}\\
 \caption{(Colour online) crystal structure of LiN$_3$ }\label{Fig 1}
 \end{figure}

\section{COMPUTATIONAL DETAILS}
 The first-principles density functional theory calculations were performed with the Cambridge Sequential Total Energy package program \cite{Payne, Segall}, using Vanderbilt-type ultrasoft pseudo potentials \cite{Vanderbilt} and a plane wave expansion of the wave functions. The electronic wave functions were obtained using  density mixing scheme \cite{Kresse} and the structures were relaxed using the Broyden, Fletcher, Goldfarb, and Shannon (BFGS) method \cite{Fischer}. The exchange-correlation potential of Ceperley and Alder \cite{Ceperley} parameterized by Perdew and Zunger \cite{PPerdew} in the local density approximation (LDA) and also the generalized gradient approximation (GGA) with the Perdew-Burke-Ernzerhof (PBE) parameterization \cite{Perdew} was used to describe the exchange-correlation potential. The pseudo atomic calculations were performed for Li 2$s^1$, N 2$s^2$ 2$p^3$. The Monkhorst-Pack scheme k-point sampling was used for integration over the Brillouin zone \cite{Monkhorst}. The convergence criteria for structure optimization and energy calculation were set to ultra fine quality. It is well known that the cut-off energy and k-point mesh influences the convergence of calculations, we tested the dependence of energy cut-off and k-point grid and found that for 520 eV plane wave cut-off energy and 5x8x5 k-point mesh, the change in total energy is less than 1meV. So we chose these plane wave cut-off energy and k-point mesh for all the calculations.
\paragraph*{}We performed the calculation by adopting the experimental crystal structure \cite{Pringle}, a=5.627$\AA$, b=3.319$\AA$, c=4.979$\AA$, and $\beta$=107.4$^0$ as the initial structure and it is relaxed to allow the ionic configurations, cell shape, and volume to change at ambient pressure. Our calculations were conducted on one unit cell with two molecules. Starting from the optimized crystal structure of lithium azide at ambient pressure, we applied hydrostatic pressure up to 60 GPa. The external pressure was gradually increased by an increment of 1 GPa in each time. Under a given pressure, the internal co-ordinates and unit cell parameters of the lithium azide crystal were determined by minimizing the Hellmann-Feynmann force on the atoms and the stress on the unit cell simultaneously. In the geometry relaxation, the self-consistent convergence on the total energy is 5x10$^{-7}$ eV/atom and the maximum force on the atom is found to be 10$^{-4}$ eV/$\AA$. Based on the equilibrium structures, the electronic, optical and vibrational properties have been calculated. The vibrational frequencies have been calculated from the response to small atomic displacements \cite{Gonze, Tulip}. The elastic constants are calculated for the optimized crystal structure at ambient conditions by using volume-conserving strain technique \cite{Mehl} as implemented in CASTEP code. We have relaxed the internal co-ordinates of the strained unit cell to arrive at the elastic constants.
%
\section{RESULTS AND DISCUSSION}
\subsection{Crystal structure and properties at ambient pressure}
At ambient pressure, LiN$_3$ crystallizes in the monoclinic structure with the C2/m space group and contains two molecules per unit cell. Each azide ion is surrounded by six cations and vice versa. The structure is iso-structural to low temperature phase of sodium azide ($\alpha$ NaN$_3$). In order to determine the theoretical equilibrium crystal structure for lithium azide, we performed a full geometry optimization of both the lattice constants and the internal atomic co-ordinates within LDA and GGA. The crystal structure of LiN$_3$ from our geometry optimization in GGA is shown in Fig 1. In Table I, we compare the lattice constants and unit cell volume with experimental data \cite{Pringle} and previous theoretical results \cite{Perger, Zhu}.
 Compared to the experimentally measured lattice constants of LiN$_3$, our LDA calculations underestimate a, b, and c by 5.3$\%$, 3.6$\%$, and 5.1$\%$ whereas GGA calculations overestimate a, b, and c by 2.3$\%$, 1.7$\%$, and 2.3$\%$, respectively. This discrepancy between theory and experiment can be expected for a molecular crystal like LiN$_3$ where the Vanderwaals forces are important which could not dealt with the LDA and GGA in the DFT calculations\cite{Chevary}. However, our calculated volume of LiN$_3$ using GGA is closer (less error) to the experimental volume when compared to LDA and therefore for further calculations we adopted GGA.
\begin{table}[tb]
\caption{
The calculated ground state properties of monoclinic LiN$_3$ at ambient pressure}
\begin{ruledtabular}
\begin{tabular}{cccccc}
  & Present work & &Other calculations & Experiment$^c$ &  \\ \hline

Lattice parameter&LDA & GGA      &    &      \\
a (\AA)     & 5.328 &5.761 &   5.696$^a$, 5.626$^b$&  5.627  \\
b (\AA)     &3.199  &3.376 &   3.235$^a$, 3.317$^b$ & 3.319 \\
c (\AA)    & 4.727 &5.094 &   5.263$^a$, 5.035$^b$&  4.979  \\
$\beta$     & 102.62 &108.6$^0$ &   113.5$^0$$^a$, 106.7$^0$$^b$ & 107.4$^0$ \\
Li   & (0, 0, 0)&(0, 0, 0) & (0, 0, 0)$^b$&\\
N   &(0.1244, 0.5, 0.7456) &(0.1005, 0.5, 0.7490) & (0.1040, 0.5, 0.7376)$^b$& (0.1048, 0.5, 0.7397)\\
N   & (0, 0.5, 0.5)&(0, 0.5, 0.5) & (0, 0.5, 0.5)$^b$& (0, 0.5, 0.5)\\
B$_0$(GPa)& 42.27 &23.23& --& 20.5$^d$
\end{tabular}
\end{ruledtabular}
$^a$ Ref (26), $^b$ Ref (8), $^c$ Ref (22), $^d$ Ref (12)
\end{table}
\subsection{Structural and electronic properties under pressure}
Starting from the zero-pressure equilibrium lattice structure, we applied hydrostatic compression to the LiN$_3$ unit cell in the pressure range from 0 to 60 GPa. This was done through CASTEP using variable cell optimization under the constraint of a diagonal stress tensor with fixed values of diagonal matrix elements equal to a desirable pressure. The obtained pressure-volume relation of LiN$_3$  is shown in Fig 2(a), together with the results of experiment by Medvedev et al\cite{Medvedev}. Our calculations show that the volume decreases monotonically with pressure and at 60 GPa the volume compression is V/V$_0$ = 57$\%$. Qualitatively our calculations reproduce the trend of pressure - induced reduction of volume by experiment (where the volume compression at 62 GPa is V/V$_0$ = 58$\%$). From the calculated P-V relation we can also observe that at a given pressure, the theoretical unit cell volume is greater than that of the experiment, which is also due to the inherent limitation of DFT-GGA functionals. However, as pressure increases, the computed cell volume approaches the experimental values. For example, if we observe keenly, the discrepancy between theoretical and experimental volumes decreases from 5.8$\%$ at 0 GPa to 2.68$\%$ at 60 GPa. This result implies that our \textit{abinitio}-GGA calculations performed under high pressure might be more reliable.
 \paragraph*{}The dependence of lattice constants of LiN$_3$ on pressure is shown in Fig 2(b), where our theoretical results are compared with the experimental data \cite{Medvedev}. The experimental trend for the lattice constants upon compression is well reproduced by our GGA calculations. Among the three axes, we find the better agreement between our GGA calculations and experiments for reduction of lattice constant for b-axis, whereas the reduction of lattice constant for a and c - axes is overestimated by our calculation. The possible reason for this behavior is the theoretical values of initial zero-pressure lattice constants a and c are 2.3$\%$ larger than the experimental values (see Table 1). In Fig 2(c), we have shown the trends in the ratios of a/a$_0$, b/b$_0$, and c/c$_0$ of LiN$_3$ as function of pressure along with the experimental data \cite{Medvedev}. It can be observed that the b-axis is the most compressible which is also consistent with experimental result. The anisotropy ratios c/a and c/b with increasing pressure are shown in Fig 2(d). The c/a ratio increases from 0.884 at 0 GPa to 0.909 at 60 GPa, whereas the c/b ratio increases from 1.508 at 0 GPa to 1.566 at 60 GPa, which indicates that the compressibility of LiN$_3$ is anisotropic. This is in good agreement with the experiment \cite{Medvedev}. The pressure dependence of the monoclinic angle of LiN$_3$ is shown in Fig 2(e) where it is also compared with experiment \cite{Medvedev}. The change in monoclinic angle is more below 20 GPa whereas above this pressure it is only of 2$^0$, this implies that the shear of layers is more below 20 GPa whereas it is less pronounced at high pressures (at 60 GPa). Due to the affect of the shear of the structure the azide ions become closer to each other and therefore the interaction will be more.
 \paragraph*{} The electronic structure of solids can be characterized by means of electronic band gap.b The electronic band structure of LiN$_3$ at 0 GPa and at 60 GPa are shown in Figs 3(a) and 3(b) respectively. The band structure at 0 GPa clearly shows that the top of the valence band and the bottom of the conduction band occurs at Z-point in the Brillouin zone indicating that the material is a direct band gap material with a separation of 3.32 eV, the value slightly lower than the previous theoretical value of 3.46 eV at the PWGGA-DZP level \cite{Perger} and 3.7 eV in Ref 6. Unfortunately, as far as we know, there is no experimental value of the band gap available for LiN$_3$ to compare our theoretical band gap result. Ofcourse, we could expect that our band gap value might be lower than the experimental band gap, because it is well known that first-principles calculation of the electronic structure of semiconductors and insulators using GGA give an underestimation of the band gap values compared to experiments \cite{Mel, Jones, Errandonea}. This underestimation of the gap is mainly due to the fact that GGA suffers from artificial electron self-interaction and also lack the derivative discontinuities of the exchange-correlation potential with respect to occupation number. One way to solve the band gap problem is to apply a perturbative correction to the energy levels, as in the GW approximation \cite{Kanchana, Bheem}. However the band gap at 60 GPa is observed to be indirect with a magnitude of 2.26 eV as it is occurred between the Z and A points. Based on the equilibrium crystal structures obtained at different pressures, we calculated the band gap and examined its change under hydrostatic compression. The calculated band gaps as functions of pressure are shown in Fig 2(f). It can be seen that the band gap reduces smoothly under compression with out any significant discontinuity. But in different pressure ranges the average decrease of the band gap is different (upto 20 GPa 0.065 eV/GPa and 0.1 eV/GPa from 20 to 60). The decrease in the band gap with pressure indicates that the electrical conductivity of LiN$_3$ increases meaning that at high pressures LiN$_3$ might changes from insulator to semi conductor.
 \paragraph*{} The nature of bonding and the electronic band gap under pressure can be studied by the total and partial DOS. Moreover, lithium azide is photosensitive and undergoes decomposition under the action of suitable wave length of light and the essential step in the photo chemical decomposition involves the promotion of a valence electron from the valance band to the conduction band, therefore it would be necessary to know the information about the type of states that are present in both valence and conduction bands. For LiN$_3$, the total and partial DOS at ambient and at 60 GPa pressure are shown in Figs 4(a) and 4(b) respectively. Overall the distribution of states is same at both pressures, and there is no hybridization between the Li states and N$_3$ states at ambient as well as at 60 GPa, indicating that ionic bond is more favored in LiN$_3$. This is also observed by using the charge density distribution plots shown in Figs 5(a) and 5(b). The \textit{p} states of azide ion are dominating at the Fermi level and as pressure increases the delocalization of the azide ion states is observed, see Fig 4(b). This is also clearly observed by the charge density distribution plots shown in Figs 5(c) and 5(d). We have also shown the charge density plots of the states at the minimum of the conduction band which are also derived mostly from the \textit{p} type azide states and a very little contribution from the \textit{s} states of Li and azide ion both at 0 GPa and 60 GPa in Figs 5(e) and 5(f) respectively.
\begin{figure}[h!]
\begin{center}
\subfigure[]{
\includegraphics[scale=0.5]{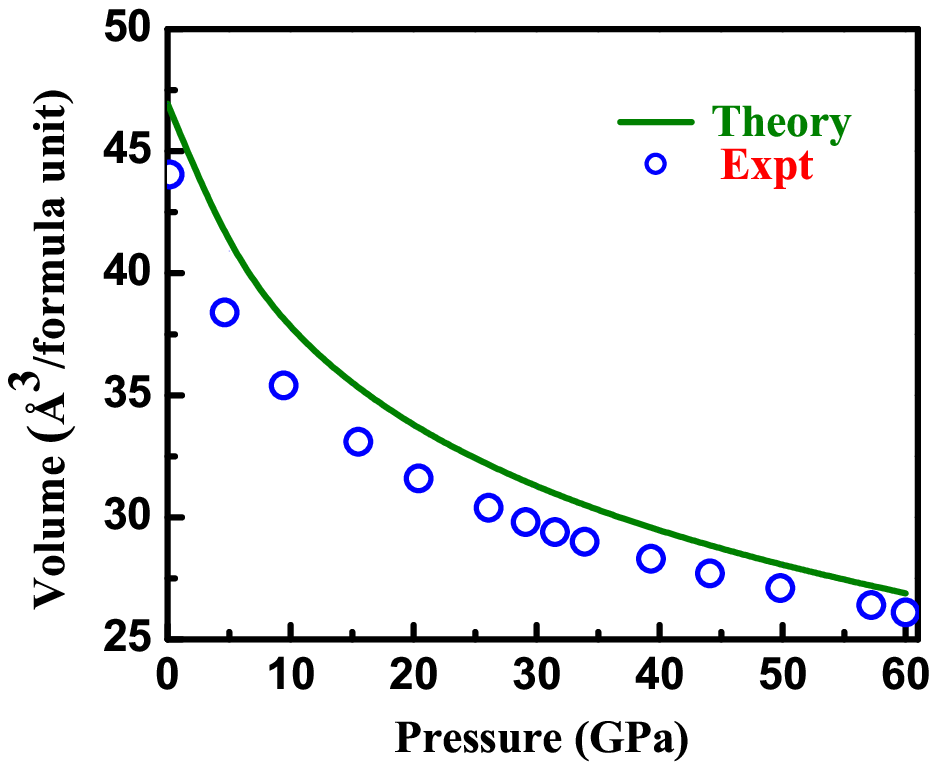}}
\subfigure[]{
\includegraphics[scale=0.5]{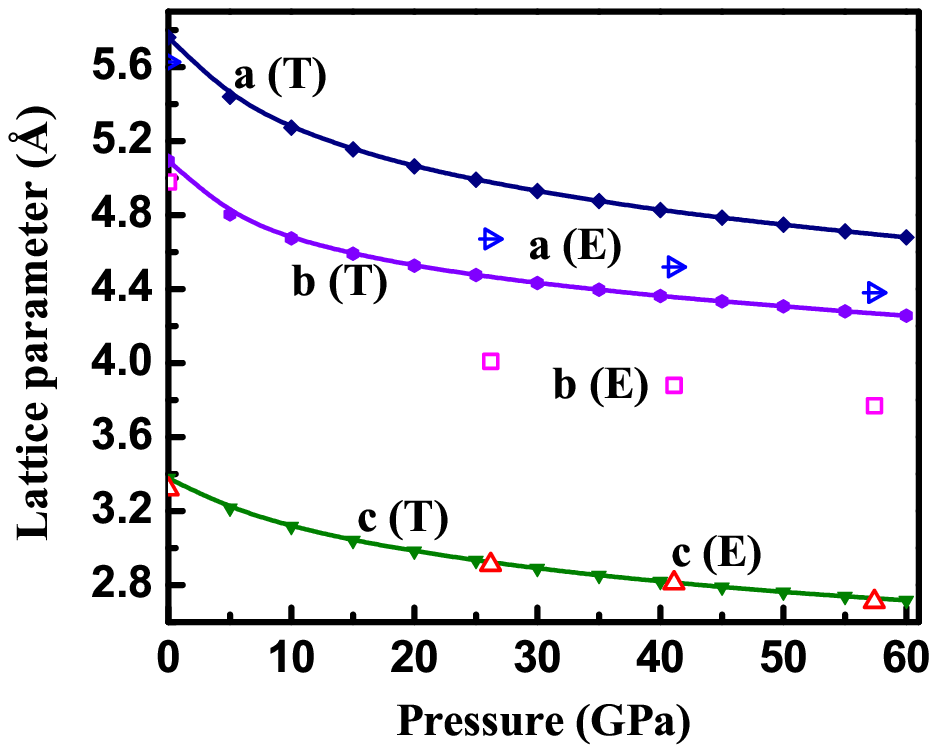}}
\subfigure[]{
\includegraphics[scale=0.5]{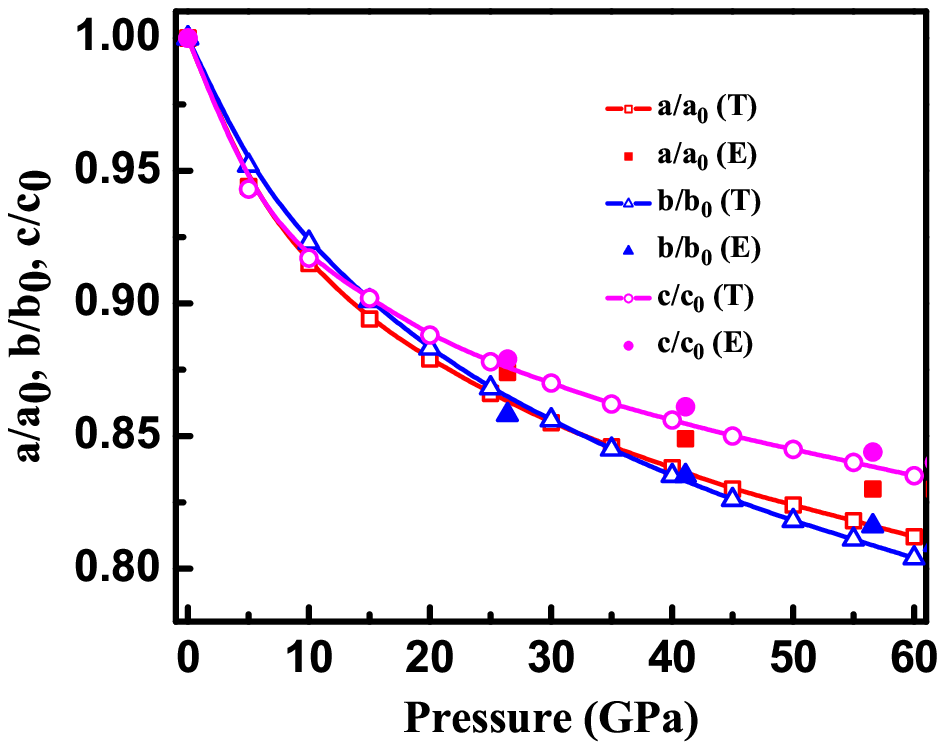}}
\subfigure[]{
\includegraphics[scale=0.5]{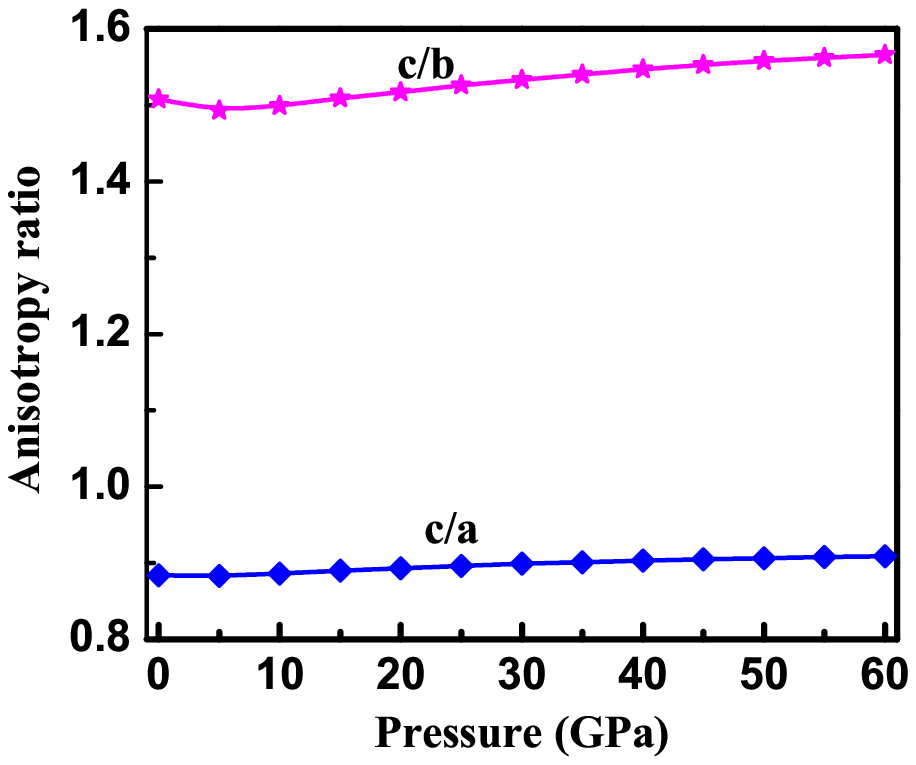}}
\subfigure[]{
\includegraphics[scale=0.5]{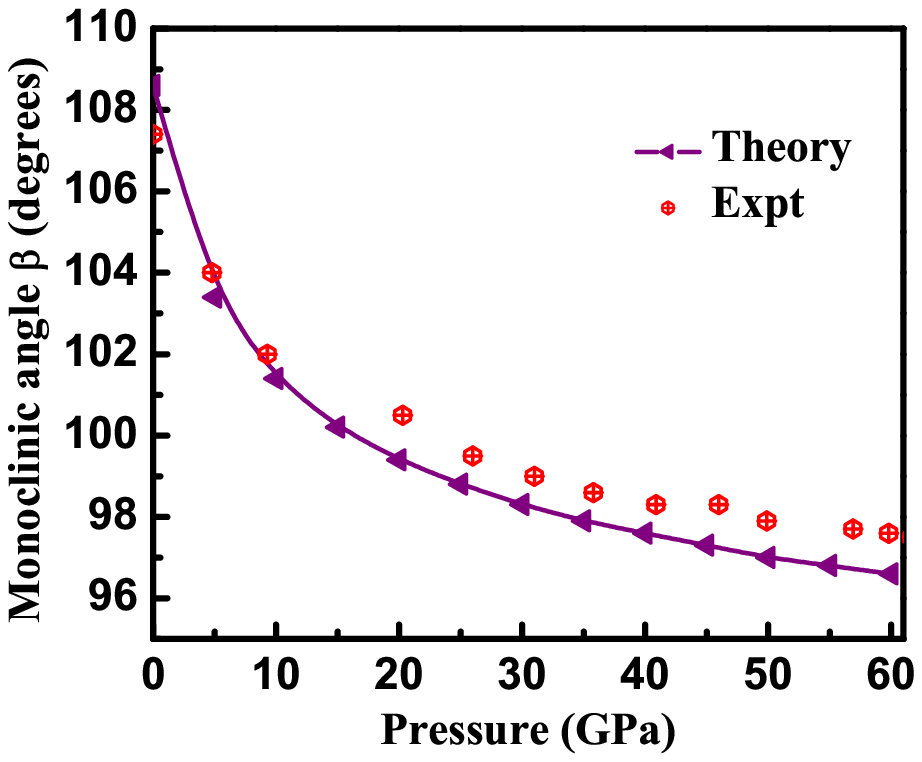}}
\subfigure[]{
\includegraphics[scale=0.5]{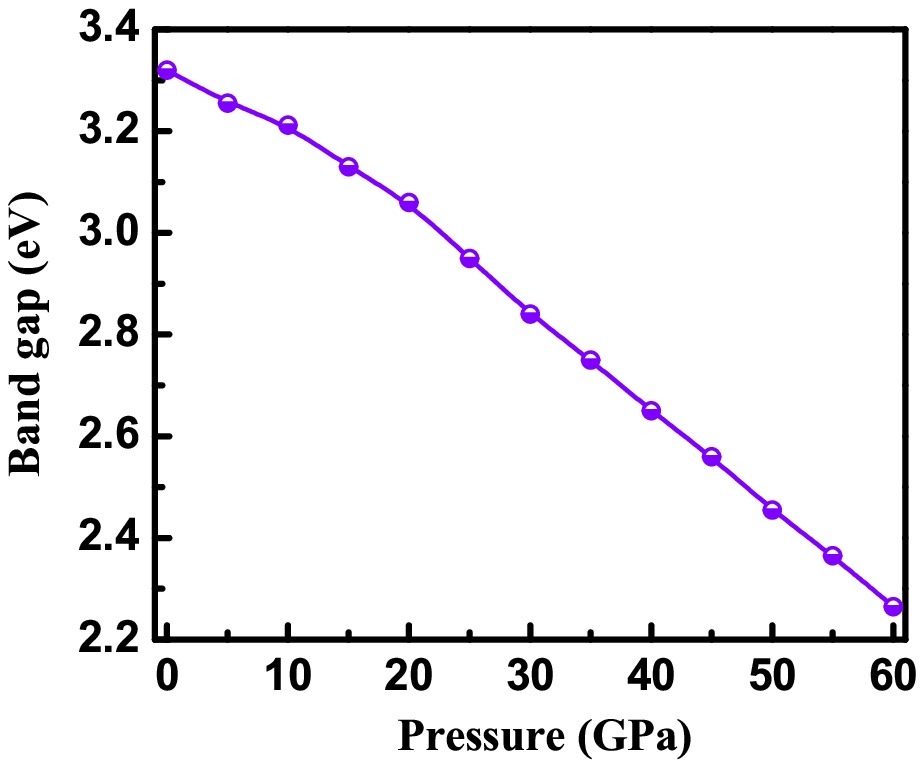}}
\caption{(Colour online) a) Volume variation of LiN$_3$ with pressure, b) Variation of lattice parameters of LiN$_3$ with pressure, c) Variation of normalized lattice parameters of LiN$_3$with pressure, d) Variation of c/a amd c/b of LiN$_3$with pressure, e) Variation of monoclinic angle $\beta$ of LiN$_3$ with pressure, f) Band gap variation of LiN$_3$ with pressure.
}
\end{center}
\end{figure}

\begin{figure}[h!]
\begin{center}
\subfigure[]{
\includegraphics[scale=1.0]{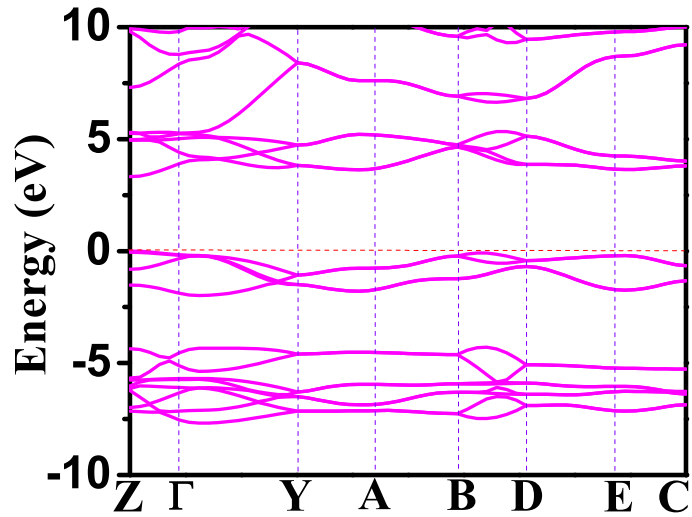}}
\subfigure[]{
\includegraphics[scale=1.0]{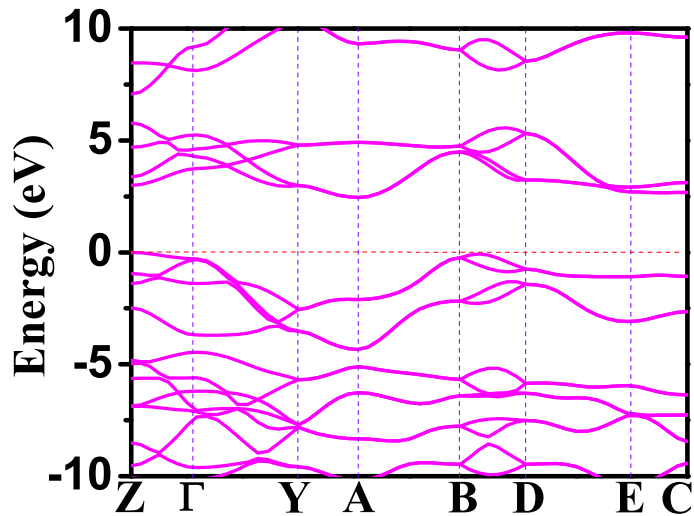}}
\caption{ (Colour online) Band structure of LiN$_3$ at 0 and 60 GPa}
\end{center}
\end{figure}

 \begin{figure}[h!]
\begin{center}
\subfigure[]{
\includegraphics[scale=1.0]{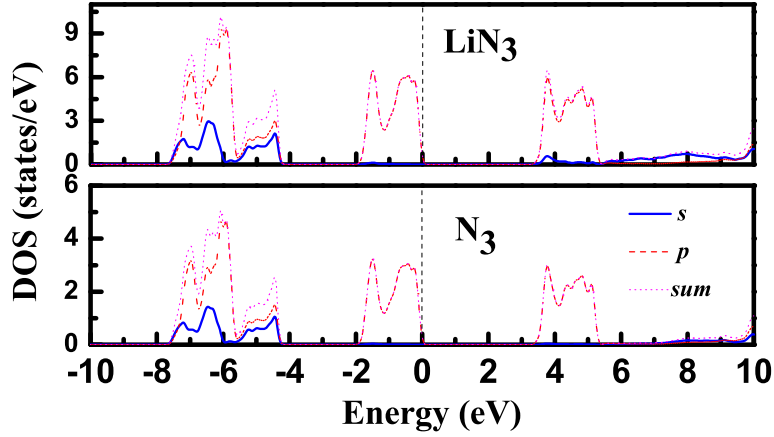}}
\subfigure[]{
\includegraphics[scale=1.0]{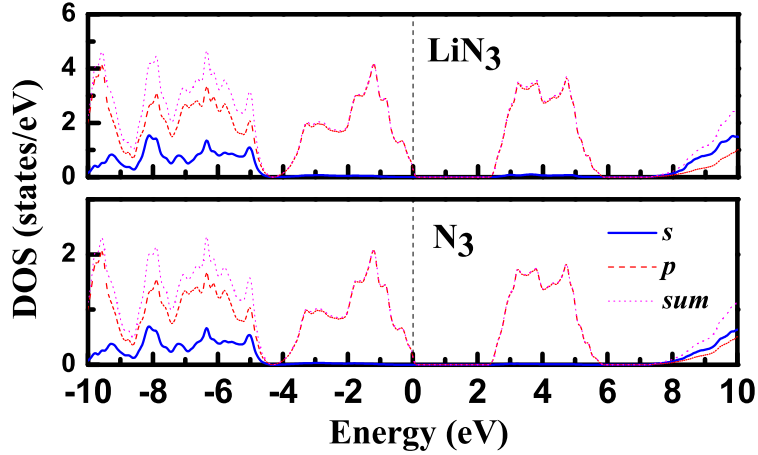}}
\caption{ (Colour online) Total and partial DOS of LiN$_3$ at 0 and 60 GPa}
\end{center}
\end{figure}

 \begin{figure}[h!]
\begin{center}
\subfigure[]{
\includegraphics[scale=0.5]{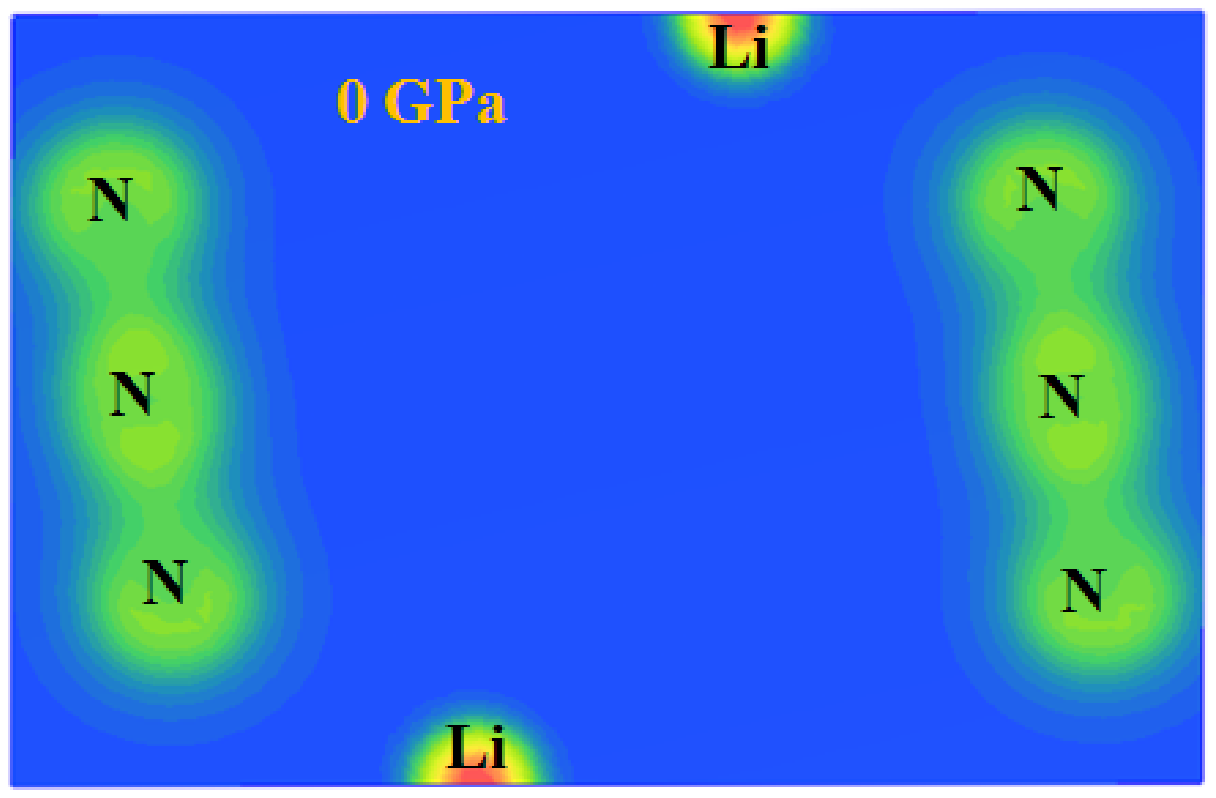}}
\subfigure[]{
\includegraphics[scale=0.4]{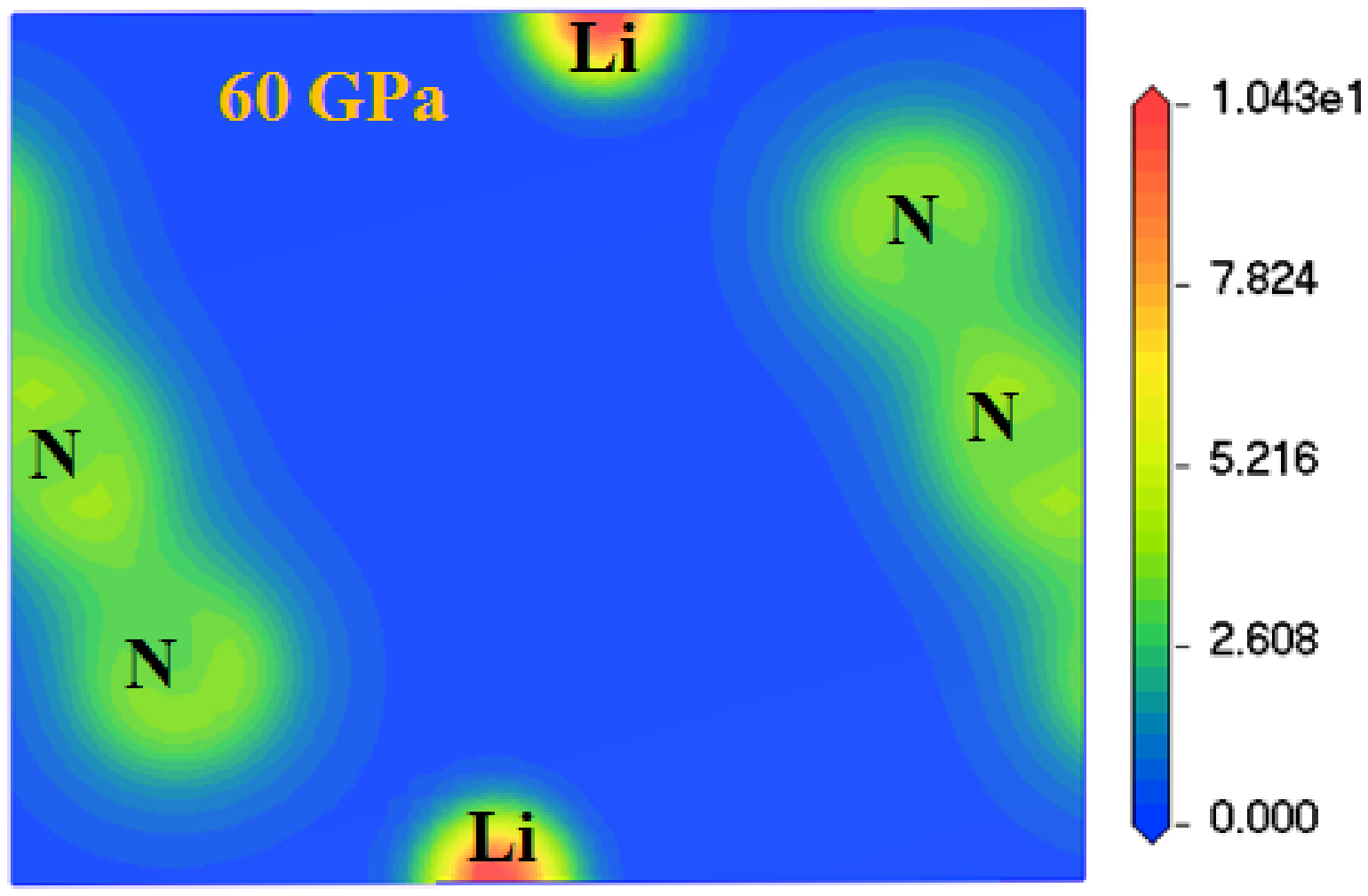}}
\subfigure[]{
\includegraphics[scale=0.4]{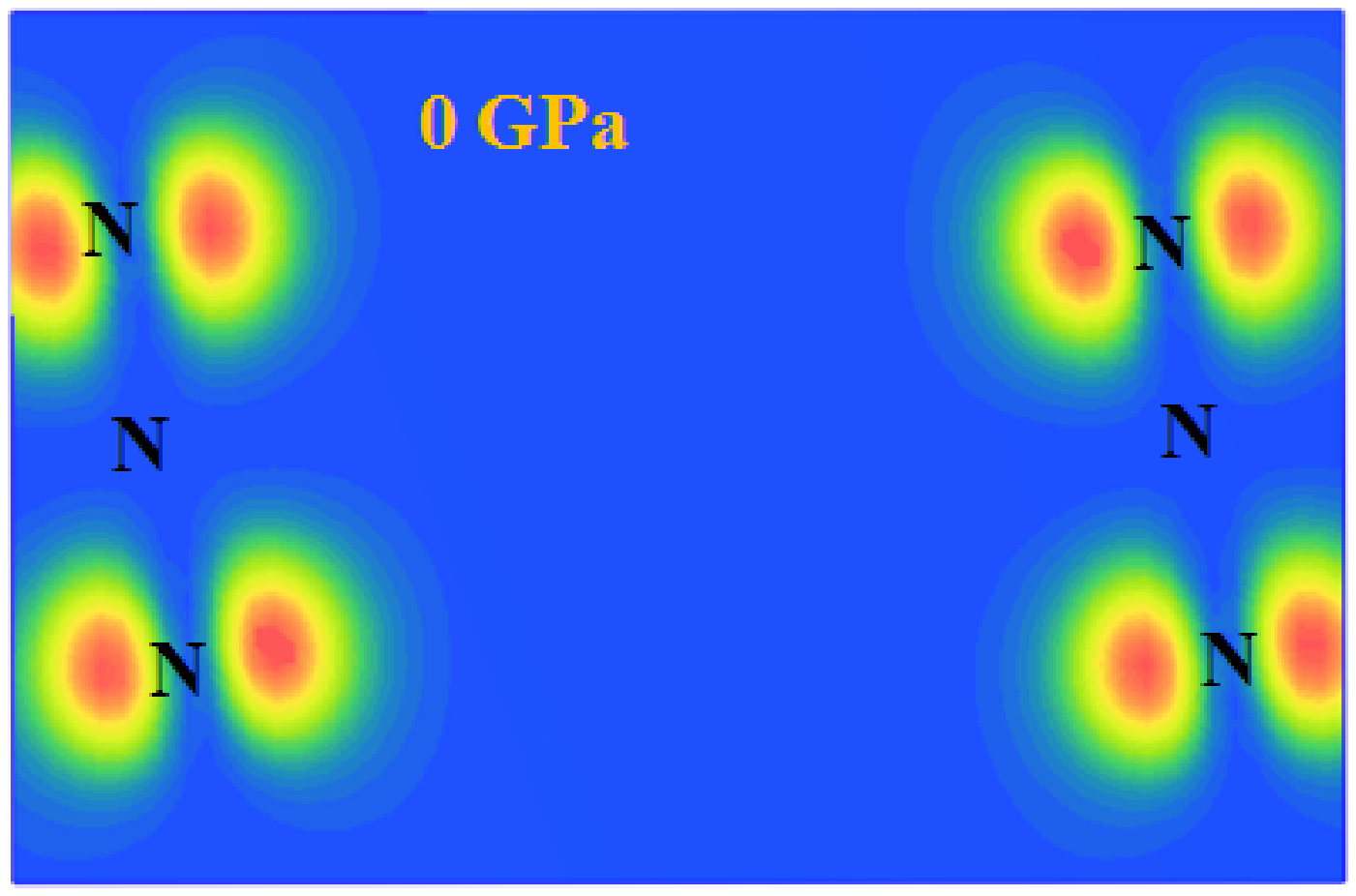}}
\subfigure[]{
\includegraphics[scale=0.4]{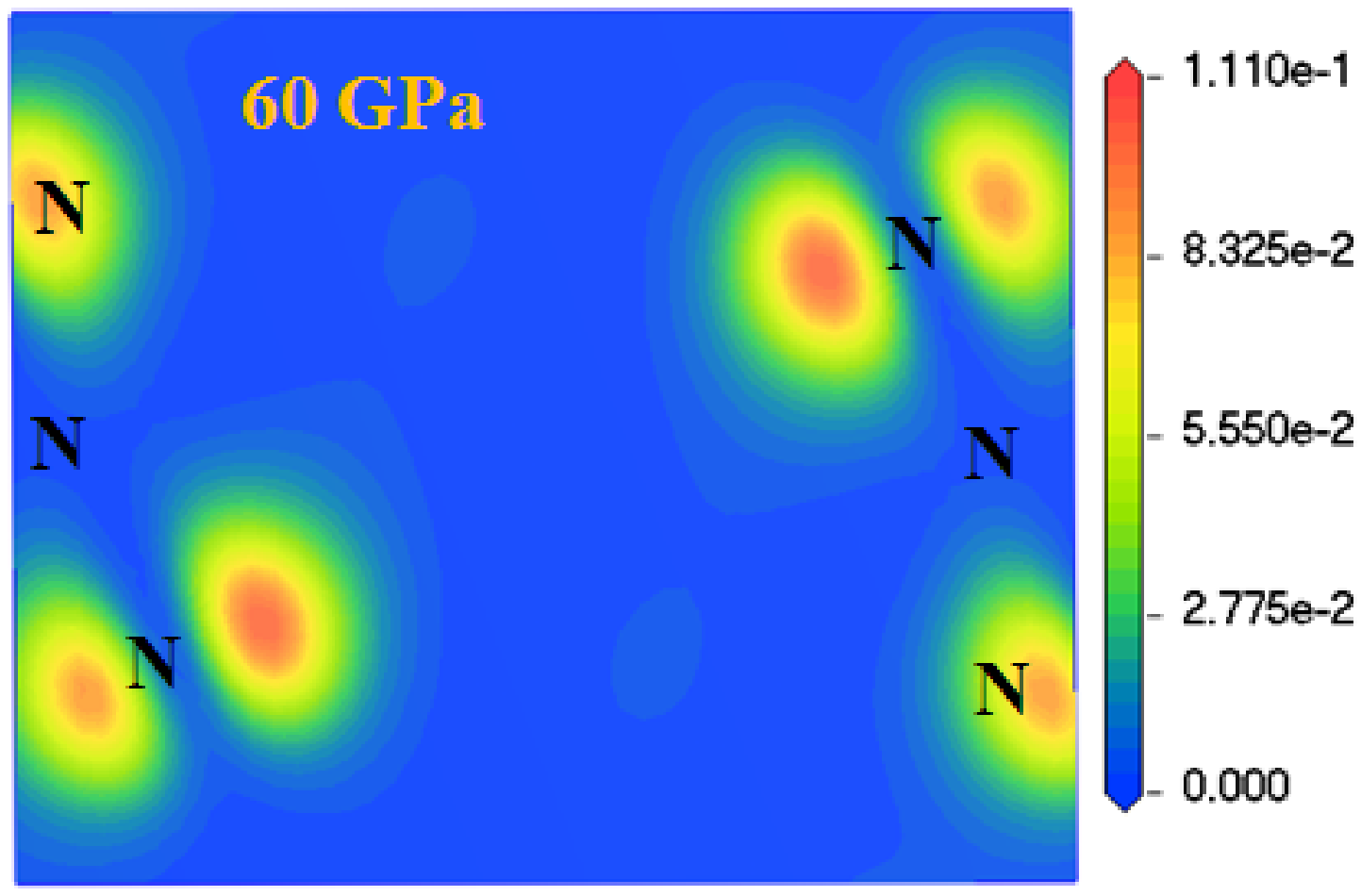}}
\subfigure[]{
\includegraphics[scale=0.4]{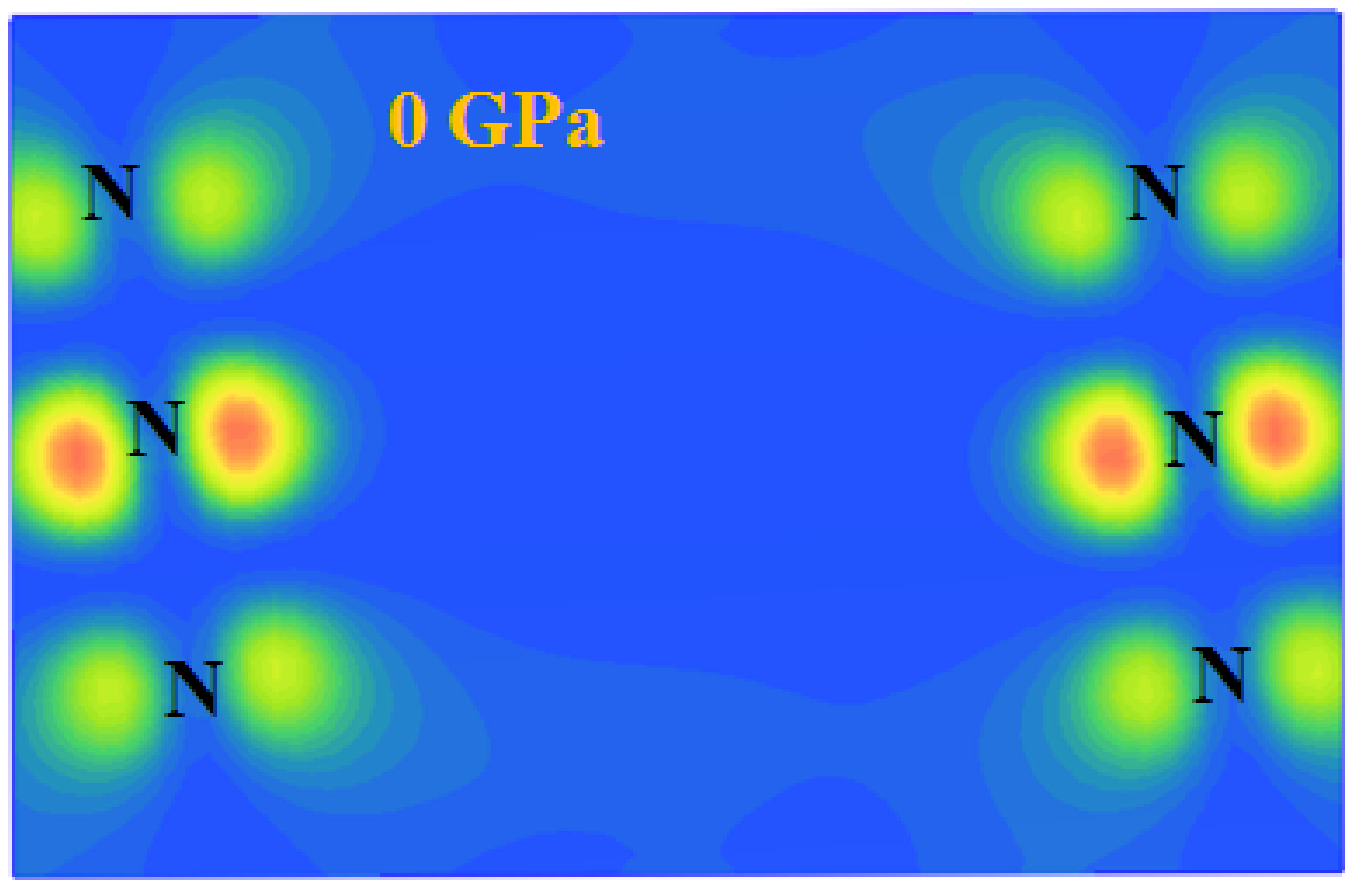}}
\subfigure[]{
\includegraphics[scale=0.4]{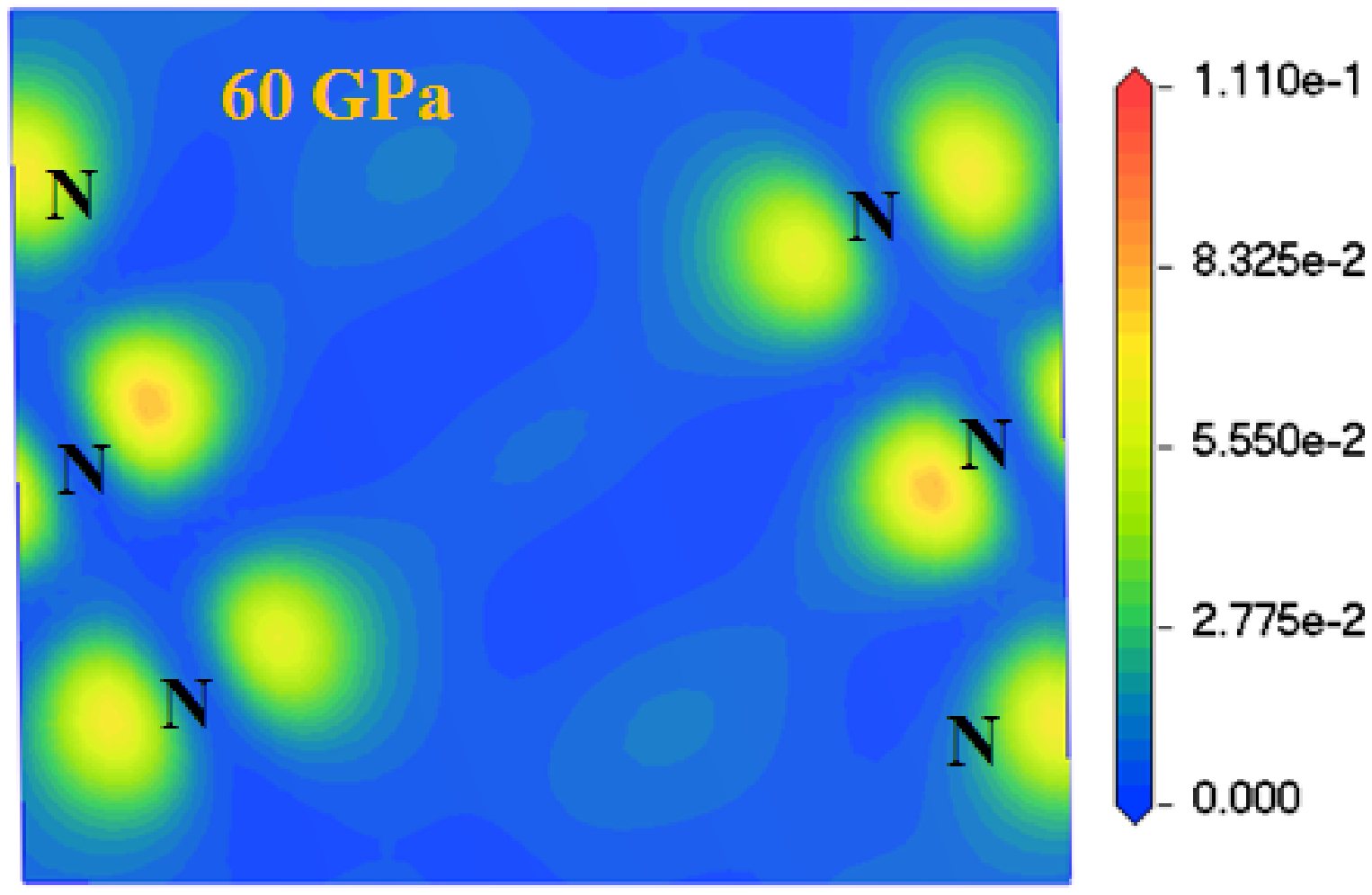}}
\caption{ (Colour online) Charge density distribution of LiN$_3$ at 0 and 60 GPa (a) and (b) are at 0 GPa and 60 GPa, (c) and (d) are of the states near by Fermi level, (e) and (f) are of minimum of the conduction band.}
\end{center}
\end{figure}

%
\subsection{Bulk modulus and elastic constants}
In order to know about the crystal stiffness and the mechanical stability at ambient conditions, we calculated the bulk modulus and elastic constants of LiN$_3$. The calculated values are listed in Table II (bulk modulus value is shown in Table I). Our calculated bulk modulus value within GGA is 23.23 GPa and it is slightly over estimated by that of the reported value of 20.5 by Medvedev et al\cite{Medvedev}, but it still comparable with that of alkali halide crystal NaCl for which B = 23.84 GPa. We have also given the calculated LDA bulk modulus value for comparison. One should note here that our LDA bulk modulus value is overestimated by that of GGA value and the experiment. This might be due to the fact that LDA overbinds the system as the volume is largely underestimated which is a common feature in DFT-LDA calculations and therefore one would expect large bulk modulus. Since the bulk modulus values are different in LDA and GGA therefore the choice of exchange-correlation functional also very important for this system. Our GGA bulk modulus value is in good agreement with that of experiment this once again shows that GGA functional is the good exchange-correlation functional for this system. As expected for ionic solids, lithium azide is a soft material because of its low bulk modulus value.  For a monoclinic crystal there are 13 independent elastic constants namely, C$_{11}$, C$_{22}$, C$_{33}$,  C$_{44}$, C$_{55}$,  C$_{66}$, C$_{12}$, C$_{13}$, C$_{15}$, C$_{23}$, C$_{25}$, C$_{35}$, and C$_{46}$.
The mechanical stability of the crystal requires the whole set of elastic constants satisfies the Born-Huang criterion \cite{Born}. 
For the case of LiN$_3$, it is found that the calculated elastic constants obey this stability criteria and therefore the monoclinic Lithium azide is mechanically a stable compound. Furthermore for a molecular crystal like LiN$_3$ one can correlate the elastic constant values with the structural properties of the crystal. Since LiN$_3$ is a monoclinic system, the elastic constants C$_{11}$, C$_{22}$, and C$_{33}$ can be directly relate to the crystallographic a, b, and c-axes respectively. The calculated values of these three constants follows the order  C$_{33}$ $>$ C$_{11}$ $>$ C$_{22}$, which implies that the elastic constant C$_{22}$ is the weakest for LiN$_3$. This reveals the fact that a relative weakness of lattice interactions is present along the crystallographic `b' axis. For the relative magnitude of the elastic constants C$_{11}$ and C$_{33}$ the analysis of crystal structure of LiN$_3$ shows that the largest number of close-contact (less than 2.3 \AA) inter molecular interactions to be situated along the crystallographic c-axis. Therefore the increased number of interactions along the c-axis would stiffen the lattice in this direction which is consistent with the value of C$_{33}$. The lowest number of interactions is along the b-axis which is further supported by the lowest value of C$_{22}$.

\begin{table}[tb]
\caption{
 Single-crystal elastic constants (C$_{ij}$in GPa)) of LiN$_3$ calculated at the theoretical equilibrium volume}
\begin{ruledtabular}
\begin{tabular}{cccccccccccccc}
Parameter& C$_{11}$  & C$_{22}$ & C$_{33}$ & C$_{44}$ & C$_{55}$ & C$_{66}$ & C$_{12}$ & C$_{13}$ & C$_{15}$ & C$_{23}$ & C$_{25}$ & C$_{35}$ & C$_{46}$  \\ \hline

&56.1 & 48.1   &  102.9    & 9.6& 26.7& 15.2 & 19.6 & 27.9 & 6.1 & 19.5 & 2.6 & 41.1 & -1.7   \\

\end{tabular}
\end{ruledtabular}
\end{table}
\subsection{Optical properties under high pressure}
Lithium azide becomes unstable and undergoes decomposition by the action of light. The photo chemical decomposition of lithium azide can be understood by the optical transitions from the valance band to the conduction band. From the total and partial DOS the states lying near the Fermi level are the \textit{p} states of azide ion, therefore the decomposition of the lithium azide can be initiated by the formation of azide radicals in the following way.
\begin{eqnarray}
N_3^- + h\upsilon\rightarrow N_3^* \\
N_3^* + E \rightarrow N_3 + e^-
\end{eqnarray}
where E is the thermal energy required to dissociate the azide radical N$_3$$^*$. The combination of two positive holes will give nitrogen gas with the liberation of energy Q and the electron trapped to the metal atom sites and thereby form the metal atom
 \begin{eqnarray}
2N_3\rightarrow 3N_2 + Q \\
Li^+ + e^- \rightarrow Li
\end{eqnarray}
Therefore in order to understand the photochemical decomposition phenomena at ambient as well as at high pressures it would be necessary to understand its optical properties that are resulting from the interband transitions. In general the optical properties of matter can be described by means of the complex dielectric function $\epsilon (\omega)$ = $\epsilon_1 (\omega)$ + $i\epsilon_2 (\omega)$, where $\epsilon_1 (\omega)$ and $\epsilon_2 (\omega)$ describes the dispersive and absorptive parts of the dielectric function. Normally there are two contributions to $\epsilon(\omega)$ namely intraband and interband transitions. The contribution from intraband transitions is important only for the case of metals. The interband transitions can further be split in to direct and indirect transitions. The indirect interband transitions involve scattering of phonons. But the indirect transitions give only a small contribution to $\epsilon(\omega)$ in comparison to the direct transitions, so we neglected them in our calculations. The direct interband contribution to the absorptive or imaginary part $\epsilon_2 (\omega)$  of the dielectric function  $\epsilon(\omega)$ in the random phase approximation without allowance for local field effects can be calculated by summing all the possible transitions from the occupied and unoccupied states with fixed k-vector over the Brillouin zone and is given as \cite{Sinha}
 \begin{equation}
\epsilon_2(\omega)=\frac{Ve^2}{2\pi\hbar m^2\omega^2}\int d^3k\sum|\langle\psi_C|p|\psi_V\rangle|^2\delta(E_C-E_V-\hbar\omega)
\end{equation}
 here  $\psi_C$ and  $\psi_V$ are the wave functions in the conduction and valence bands, $p$ is the momentum operator, $\omega$  is the photon frequency, and $\hbar$  is the Planck's constant.
The real part $\epsilon_1 (\omega)$ of the dielectric function can be evaluated from $\epsilon_2 (\omega)$ using the Kramer–Kroning relations.
\begin{equation}
\epsilon_1(\omega)=1+\frac{2}{\pi}P\int_0^\infty \frac{\epsilon_2(\omega^\prime)\omega^\prime d\omega^\prime}{(\omega^\prime)^2-(\omega)^2}
\end{equation}
where `P' is the principle value of the integral.The knowledge of both the real and imaginary parts of the dielectric function allows the calculation of the important optical properties such as refractive index, absorption, and photo conductivity \cite{Sinha}.
\paragraph*{} In this present study, we investigated the static dielectric constant, refractive index, absorption spectrum and photo conductivity of lithium azide along the three crystal directions at ambient pressure as well as at the high pressures. The static dielectric constant $\epsilon_{1}(0)$ as a function of pressure is shown in Fig 6(a). Clearly $\epsilon_{1}(0)$ increases with pressure along all the three directions. The static refractive index (n=$\sqrt\epsilon(0)$)  of the system in three directions is given by n$_{100}$ = 1.73,  n$_{010}$ = 1.34, and  n$_{001}$ = 2.61. Clearly, n$_{100}$ $\neq$ n$_{010}$ $\neq$ n$_{001}$ therefore we conclude that the lithium azide is an anisotropic crystal with bi-axial crystal properties. At ambient pressure, our calculated values of $\epsilon_{1}(0)$ and n(0) along [100] direction are in good agreement with that of previous calculations \cite{Zhu}. The calculated refractive index increases with increase in pressure as shown in Fig 6(b), implies that the crystal binding will be more under the application of pressure. In Fig 6(c), we have shown the calculated absorption spectra and photo conductivity spectra of LiN$_3$ along [100], [010], and [001] directions at 0 GPa. Our calculated results indicates that there is anisotropy in the calculated optical spectra. Our calculated absorption spectra along [100] direction is in good agreement with that of the reported absorption spectra in Ref [8]. The peaks in the absorption spectra are due to the interband transitions from the occupied to unoccupied states. The absorption starts from the energy of 3.32 eV, which is the energy gap between the valance and conduction bands. The first absorption peak in all the three directions is observed at (5.98 eV) but the absorption coefficient is found to be different and it is around 1.9x10$^7$ m$^{-1}$ in [100] and [010] directions whereas it is 4.5x10$^7$ m$^{-1}$along [001] direction. From this result the decomposition of lithium azide takes place under the action of ultra violet light that starts from the absorption edge of 3.32 eV.
\paragraph*{} Photo conductivity is due to the increase in number of free carriers when photons are absorbed. The calculated photo conductivity shows that LiN$_3$ is photo sensitive and display a wide photo current response in the absorption region of 3.3 to 20 eV. The calculated absorption and photo conductivity spectra at pressures of 20 GPa, 40 GPa, 60 GPa are shown in Fig 6(d), 6(e), and 6(f) respectively. It can be clearly observe that the fundamental absorbtion energy region and also the absorption coefficient along the three crystallographic directions increases with pressure. At 0 GPa, the fundamental absorption was found to be between 3.3 to 20 eV and at 20 GPa, it increases to 23 eV and it further increases to 25 and 28 eV at 40 and 60 GPa. And also the absorption coefficient along the three directions increases with pressure over the entire absorption energy region. This means that as pressure increases the response of the electrons to the incident photon energy increases. This can be understood by the increase in photo conductivity, which shows a broad photo current response in the absorption region as pressure increases. Overall, lithium azide shows a strong absorption (absorption coefficient of the order of 10$^7$ m$^{-1}$) as a function of pressure over a broad energy range. Therefore from the study of optical properties of Lithium azide under pressure we came to a conclusion that, although LiN$_3$ is stable up to 60 GPa it may become unstable or decompose under the action of light which is equal to the absorption energy range of LiN$_3$ (ultra violet light). In the above case LiN$_3$ becomes more unstable and decomposes easily into metal and nitrogen under pressure by the influence of ultraviolet light.

\subsection{Vibrational properties}
Vibrational properties are obtained by the use of linear response method within the density functional perturbation theory (DFPT) \cite{Gonze}. In this method the force constants matrix can be obtained by differentiating the Hellmann-Feynman forces on atoms with respect to the ionic co-ordinates. This means that the force constant matrix depends on the ground state electron charge density and on its linear response to a distortion of atomic positions. By variational principle the second order change in energy depends on the first order change in the electron density and this can be obtained by minimizing the second order perturbation in energy which gives the first order changes in the density, wave functions, and potential. In the present study the dynamical matrix elements are calculated on the 5x8x5 grid of k-points using the linear response approach. The calculated total and partial phonon density of states of LiN$_3$ at 0 GPa and 60 GPa are shown in Fig 7 and Fig 8. From the partial phonon density of states it can be observed that at 0 GPa, the frequency modes below 300 cm$^{-1}$ are due to the both lithium and azide ion, whereas above this frequency the states are entirely dominated by the the azide ion. At 60 GPa, the modes are shifting towards high frequency region.

\paragraph*{} The  vibrational frequencies at gamma point are shown in  Table III. The group symmetry decomposition into irreducible representations of the C2/m point group yields a sum of A$_u$+2B$_u$ for three acoustic modes and 4A$_g$+2B$_g$+5A$_u$+10B$_u$ for the 21 optical modes. The modes from M1 to M13 involves the vibrations from the lattice (including both metal atom and azide ion) whereas modes from M14 to M21 are entirely due to the azide ion. The M14 mode, which is due to the N$_3$ symmetric bending along b-axis, has frequency 640.02 cm$^{-1}$ is in good agreement with that of the experimental value \cite{Evans} of 635 cm$^{-1}$. The M15 mode, located at 645.45 cm$^{-1}$ originates from the asymmetric bending of azide ion along b-axis. The M16 and M17 modes having frequencies 645.45 cm$^{-1}$ and 645.62 cm$^{-1}$ are due to the symmetric and asymmetric bending of azide ion along a-axis. The modes from M18 to M21 are the stretching modes of azide ion. Our calculated value of 1262.51 cm$^{-1}$ for the symmetric stretching of azide ion is in good agreement with that of the experimental value \cite{Evans} 1277 cm$^{-1}$. The asymmetric stretching frequency of azide ion is found to be 2006.51 cm$^{-1}$, the value lower than that of experimental value 2092 cm$^{-1}$. This discrepancy may be due to the overestimation of the crystal volume because of GGA exchange-correlation functionals. In order to understand the intramolecular and intermolecular interactions under compression, we calculated the vibrational frequencies for the optimized crystal structures up to the pressure of 60 GPa at a pressure step of 5 GPa and are shown in Fig 9. It can be clearly observe that as pressure increases all the vibrational frequencies of all modes from M1 to M21 increases with increase in pressure, except the bending modes of azide ion from M14 to M17 which decreases and have minima at about 30 GPa for M14 and M15 modes, at about 10 GPa for M16 mode and at 5 GPa for M17 mode. This clearly shows that the intramolecular interaction enhances under the application of pressure. In Table III we have shown the coefficients for the pressure-induced shifts of the vibrational frequencies of all vibrational modes of LiN$_3$ from M1 to M21. These values are obtained by a linear fit of the vibrational frequencies with respect to pressure. From the calculated values it is clear that different vibrational modes show distinctly different pressure dependent behavior. The pressure-coefficients for lattices modes are found to be higher than that of internal modes, indicating that the lattice modes show most significant shift over internal modes. The lattice mode M13 has high pressure coefficient of 8.1 cm$^{-1}$/GPa and among the internal modes, the asymmetric stretching modes of azide ion M20 and M21 have high pressure coefficient of 3.3 cm$^{-1}$/GPa. Over all the behavior of vibrational modes under compression indicates that the intermolecular and intramolecular interactions increases. The observed frequency shifts are more for lattice modes compared to the internal modes under compression. This implies that intermolecular interaction is affected more significantly than the intramolecular interaction under compression.

\paragraph*{}
The knowledge of the volume dependencies of the vibrational modes allows to calculate the Gr\"{u}neisen parameters ($\gamma_i$) associated with them. The Gr\"{u}neisen parameter of i$^{th}$ vibrational mode can be defined as
\begin{equation}
\gamma_i = -\frac{\partial{(ln \nu_i)}}{\partial(ln V)}
\end{equation}
The calculated Gr\"{u}neisen parameters ($\gamma_i$) of LiN$_3$ resulting from a linear fit of the ln $\nu$ as a function of ln V are listed in Table III. It can be observed from Table III that the calculated $\gamma$ is high for lattice modes and the values are low for internal modes which are entirely due to the azide ion vibrations. This also implies that the change in lattice parameters and thereby the volume with pressure has large affect on vibrational modes especially lattice modes whereas the vibrational frequencies of azide ion has less influence by the pressure.
\begin{table}[tb]
\caption{
The calculated vibrational frequencies of monoclinic LiN$_3$ at ambient pressure, Pressure coefficients (cm$^{-1}$/GPa) and Gr\"{u}neisen parameters ($\gamma$) of the vibrational frequencies of all vibrational modes of monoclinic LiN$_3$ from the linear fit of the present GGA results. Values in parenthesis are from experiment}
\begin{ruledtabular}
\begin{tabular}{cccccc}
 Mode & Symmetry &Frequency (cm$^{-1}$) & Pressure coefficient(cm$^{-1}$/GPa)& Gruneisen Parameter ($\gamma$)& \\ \hline
M1 & B$_u$  &138& 2.1&1.26 \\
M2 & A$_u$  & 140.2& 2.3&1.31 \\
M3 & A$_u$  &143.4& 3.4& 1.54 \\
M4 & B$_u$  &155.9&3.4 & 1.52  \\
M5 & A$_u$  &172.5&3.1&1.40 \\
M6 & B$_u$  &  189.7& 4.6&1.66  \\
M7 & B$_g$  &191.6 &4.9&1.68 \\
M8&  B$_u$  &202.3&5.7&1.87\\
M9 & B$_u$  &210.1&5.5&1.79 \\
M10 & A$_g$  &212.2&5.3&1.71 \\
M11 & B$_u$  &220.7&6.2&1.84\\
M12 & B$_u$  &264.6&5.8&1.56 \\
M13 & A$_g$  &304.6 &8.1&1.53 \\
M14 & A$_u$  &640.0 (635)$^a$&0.5&0.04  \\
M15 & B$_u$  &645.4&0.5&0.03 \\
M16& A$_u$  &645.6&1.1&0.14 \\
M17 & B$_u$  &661.3&0.8&0.11\\
M18 & A$_g$  &1262.5&1.6&0.14\\
M19 & A$_g$  &1269.5(1277)$^a$&2.2&0.21\\
M20 & B$_u$  &   1933.3&3.3&0.19\\
M21 & B$_u$  &2006.5(2092)$^a$&3.3&0.17\\
\end{tabular}
\end{ruledtabular}
$^a$ Ref (3).
\end{table}

\begin{figure}[h!]
\begin{center}
\subfigure[]{
\includegraphics[scale=0.5]{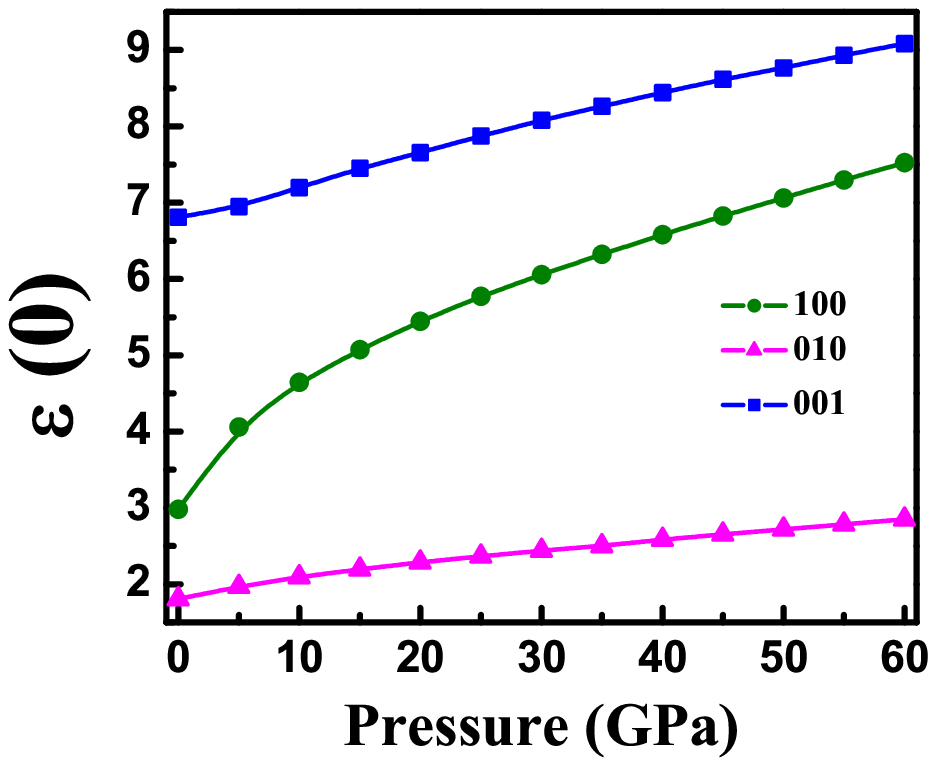}}
\subfigure[]{
\includegraphics[scale=0.5]{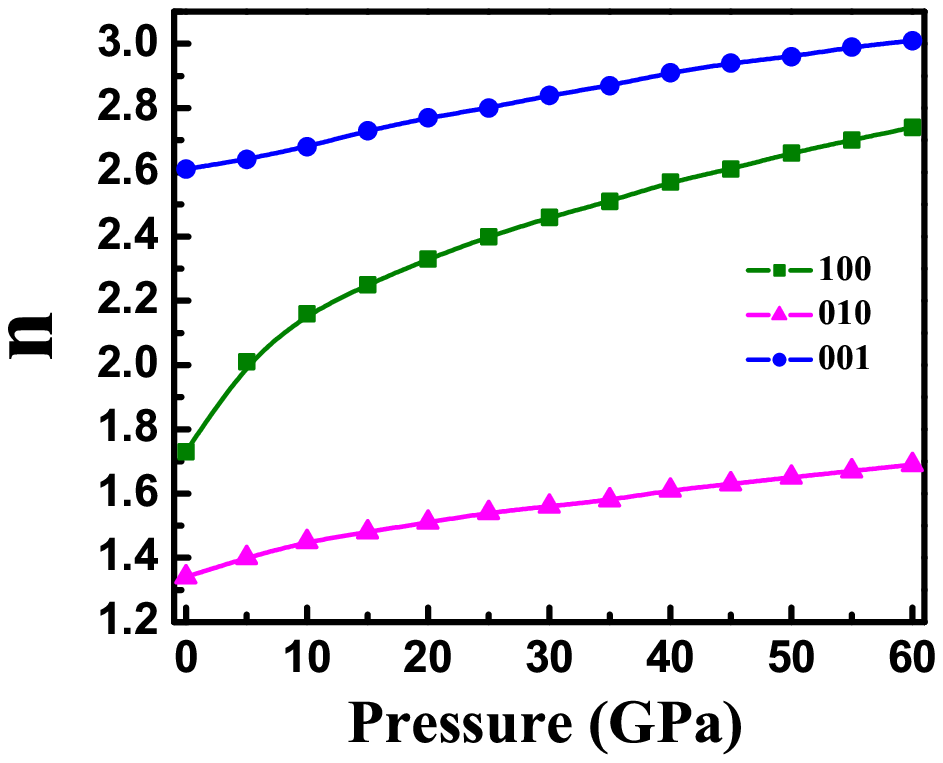}}
\subfigure[]{
\includegraphics[scale=0.5]{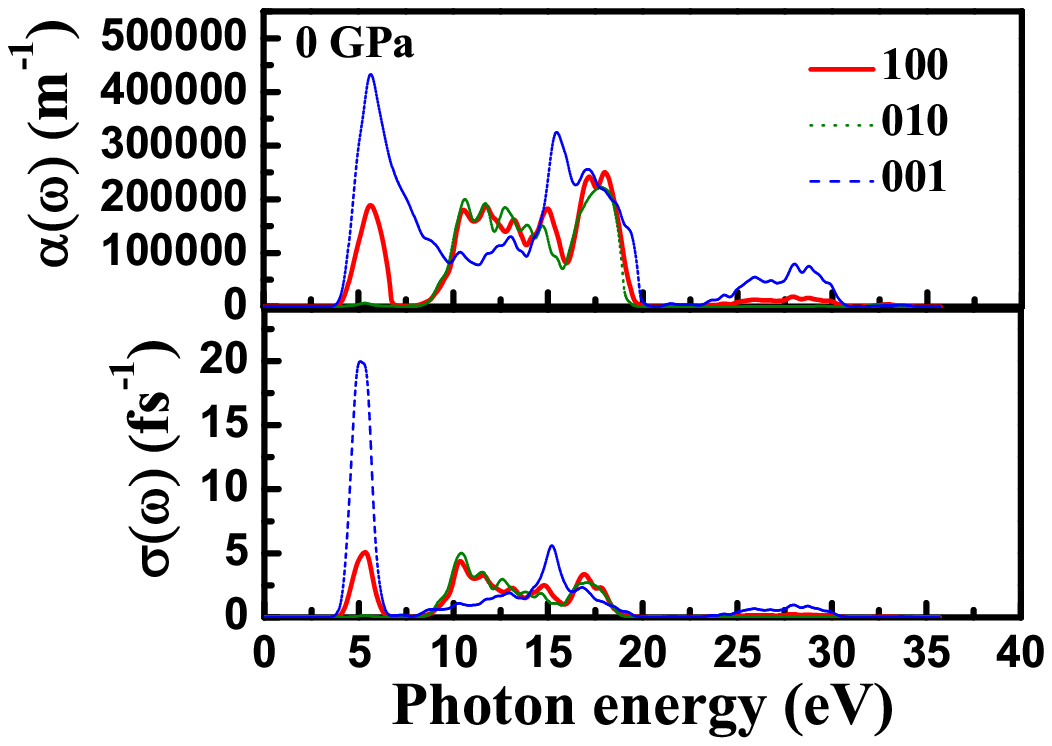}}
\subfigure[]{
\includegraphics[scale=0.5]{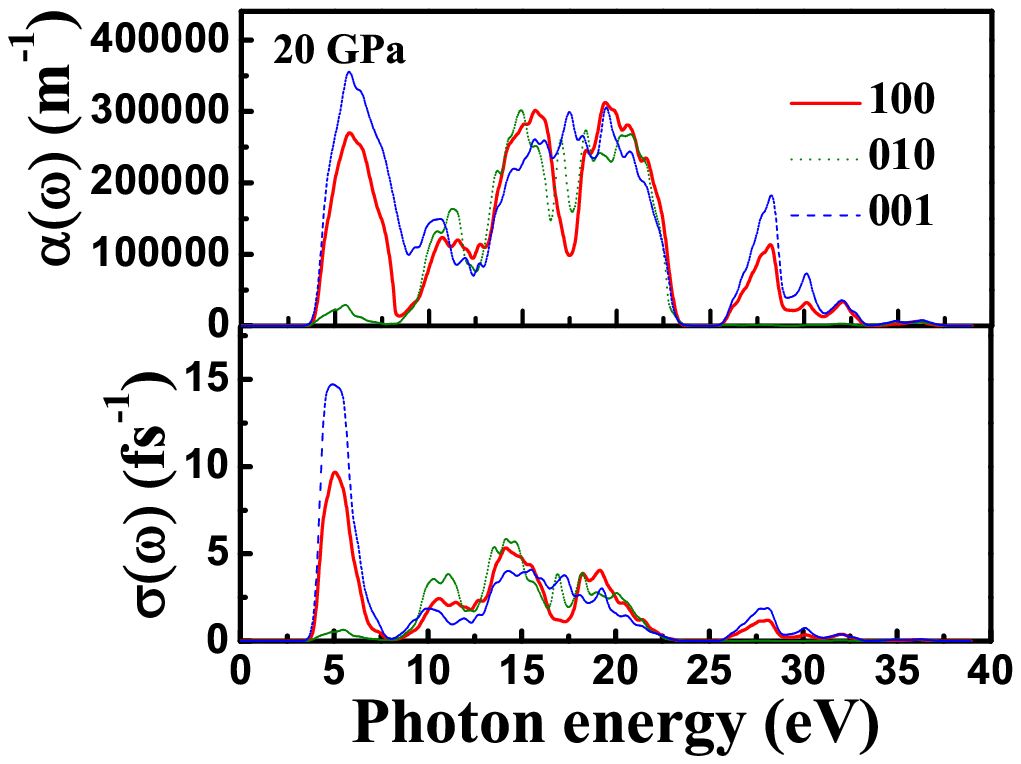}}
\subfigure[]{
\includegraphics[scale=0.5]{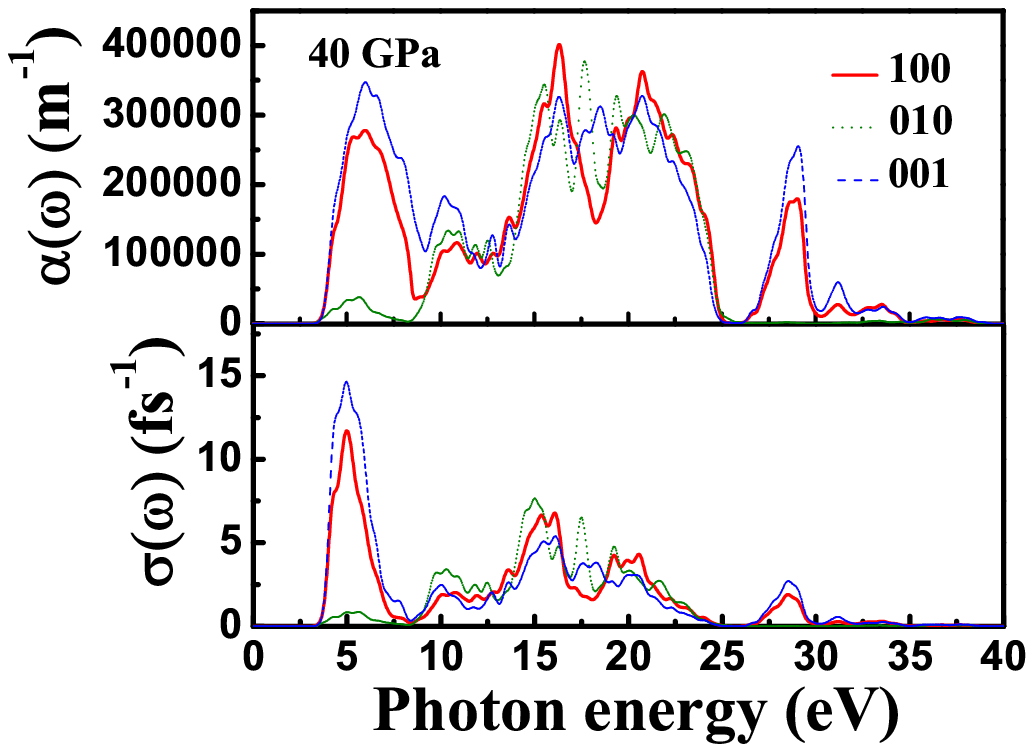}}
\subfigure[]{
\includegraphics[scale=0.5]{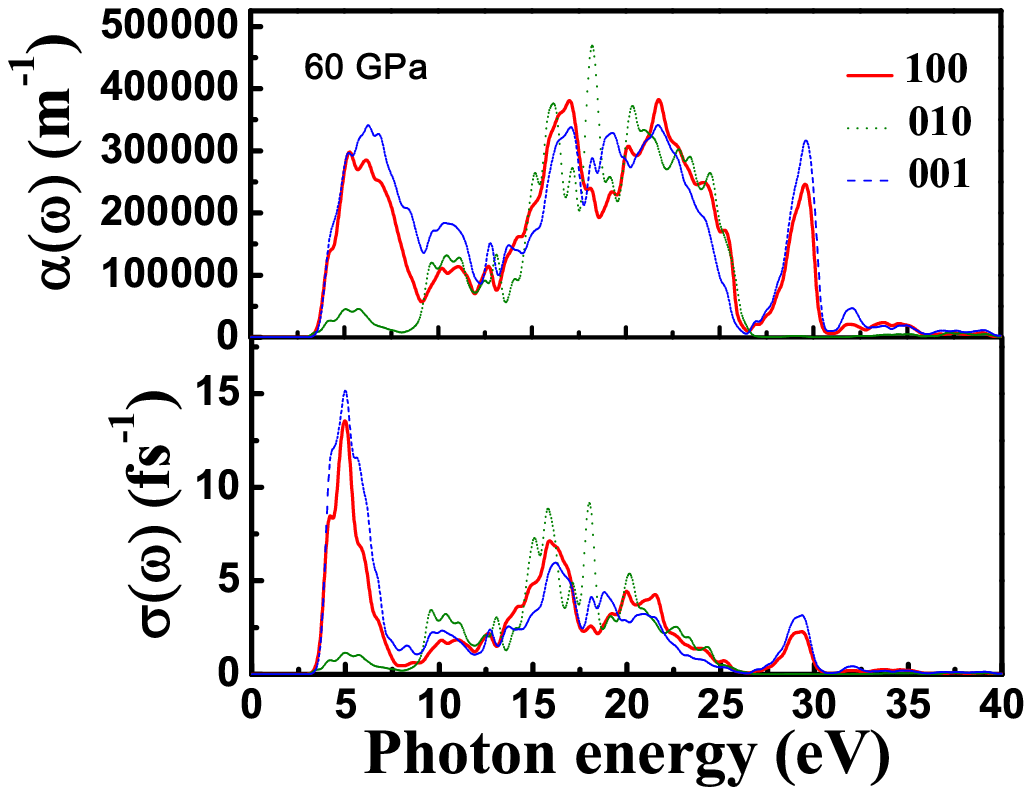}}
\caption{(Colour online) (a) The calculated static dielectric constant as a function of pressure. (b) The calculated refractive index of LiN$_3$ as a function of pressure. (c) The calculated absorption and photo conductivity of LiN$_3$ at the pressure of 0 GPa (d) at 20 GPa (e) at 40 GPa and (f) at 60 GPa.}\label{Fig 6}.
\end{center}
\end{figure}

\begin{figure}
\centering
\includegraphics[scale=1.00]{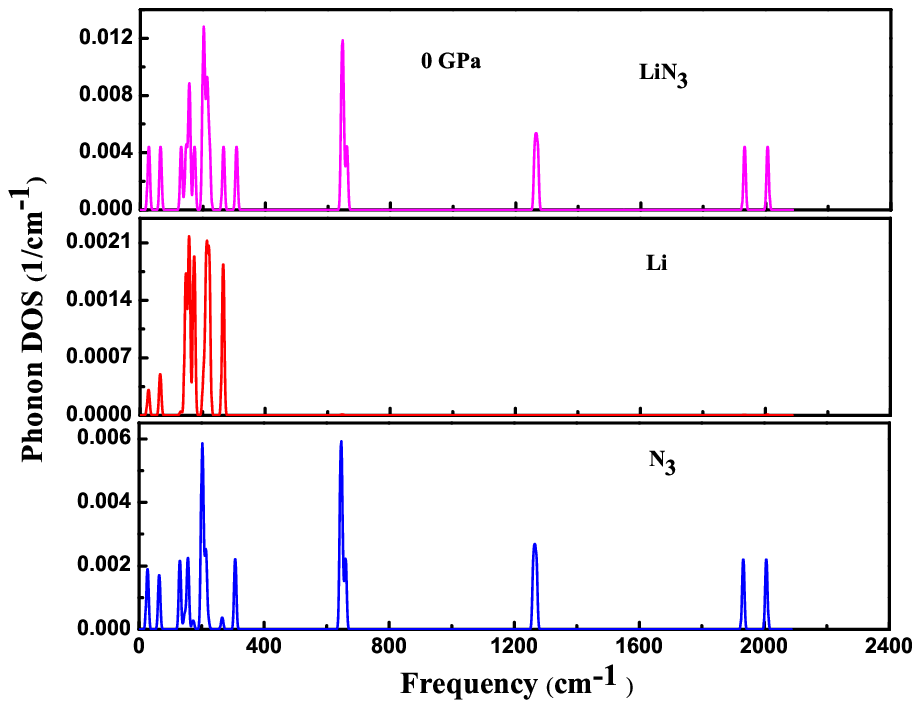}\\
 \caption{(Colour online) Total and partial Phonon density of states (PhDOS) of LiN$_3$ at 0 GPa} \label{Fig 1}
 \end{figure}

 \begin{figure}
\centering
\includegraphics[scale=1.00]{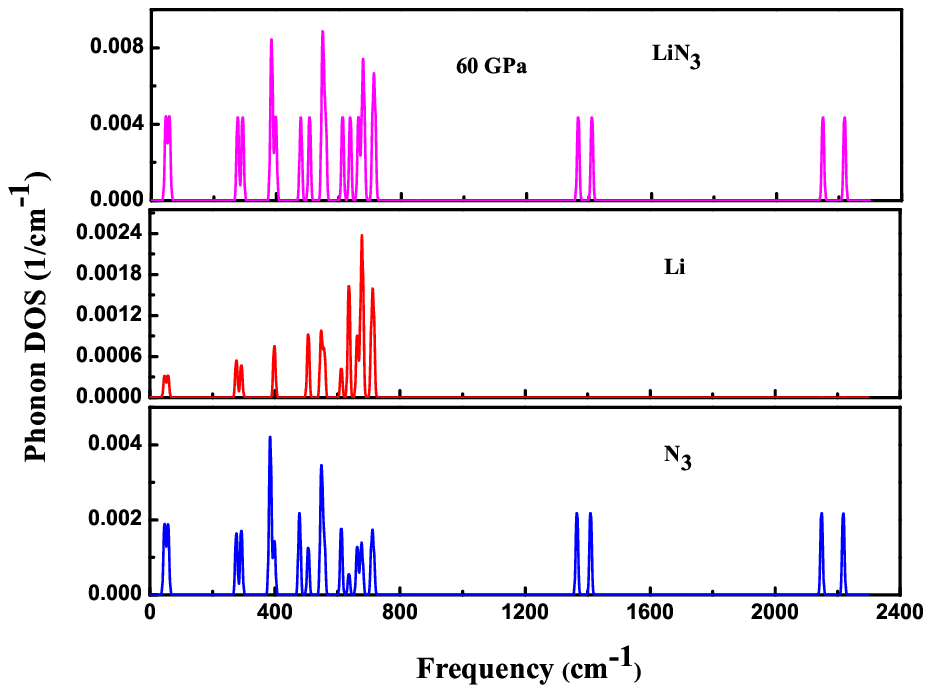}\\
 \caption{(Colour online) Total and partial Phonon density of states (PhDOS) of LiN$_3$ at 60 GPa}\label{Fig 1}
 \end{figure}
\begin{figure}
\centering
\includegraphics[scale=1.50]{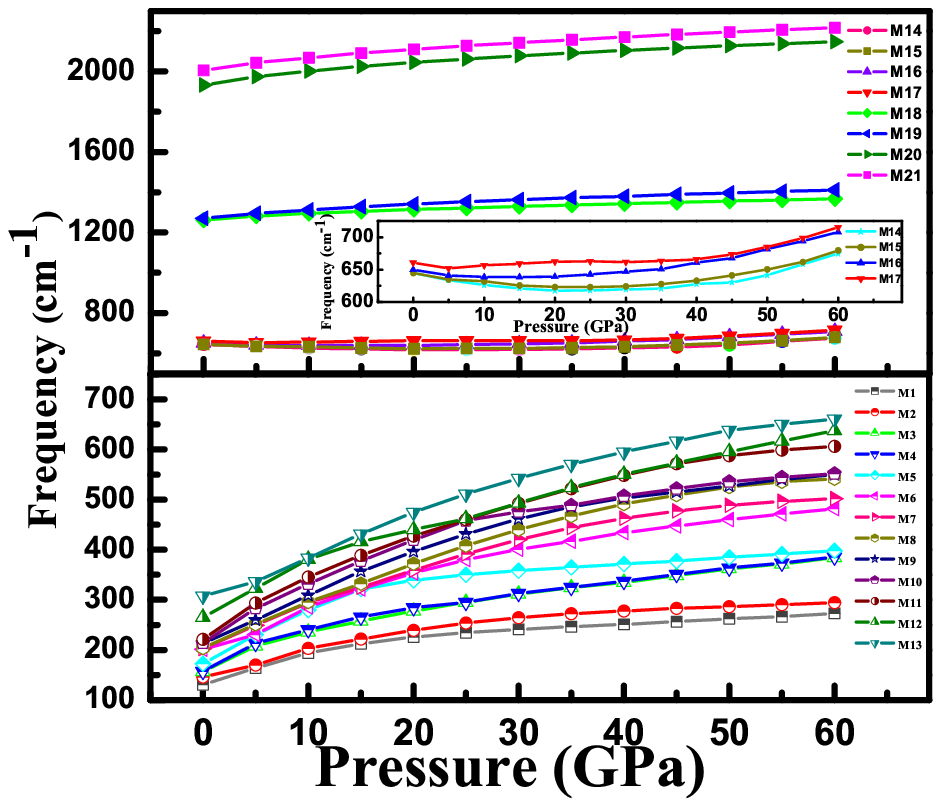}\\
 \caption{(Colour online) Pressure-induced shifts of vibrational frequencies of LiN$_3$. The inset figure shows the enlarged case of pressure dependence of M14 to M17 modes.}\label{Fig 1}
 \end{figure}

\section{CONCLUSIONS}
In summary, the structural, electronic, optical and vibrational properties of lithium azide under hydrostatic compression up to 60 GPa have been studied using density functional theory within generalized gradient approximation. The calculated structural parameters are overestimated compared to the experiment, which is due to the GGA exchange-correlation functionals. It is also observed that Lithium azide remains in the monoclinic structure in the studied pressure range of 60 GPa as observed in experiment. From the calculation of total and partial DOS it is found that lithium azide is an ionic solid with a band gap of 3.32 eV and the gap decreases as pressure increases indicates its ability to become semiconductor at high pressures. The single crystal elastic constants at ambient pressure have been calculated and found that the system is mechanically stable. The calculation of refractive index suggests that the LiN$_3$ is anisotropic material and the anisotropy increases with pressure. The absorption spectra and photo conductivity spectra at various pressures have been calculated and found that the absorption peaks are shifting towards the high energy region and photo current increases as pressure increases. Therefore we conclude that the decomposition of lithium azide is more favorable by the action of light (ultra violet) under pressure. The vibrational frequencies at ambient and high pressures have been calculated. At ambient pressure the calculated vibrational frequencies are in good agreement with the experiment values. The calculated vibrational frequencies at high pressures together with pressure coefficients and Gr\"{u}neisen parameters show that the intermolecular interactions are highly affected by the applied pressure.

 \section{ACKNOWLEDGMENTS}
 K R B would like to thank DRDO through ACRHEM for financial support. All the authors acknowledge CMSD, University of Hyderabad for computational facilities.

*Author for Correspondence,
E-mail: gvsp@uohyd.ernet.in

\clearpage
\newpage

\subsection*{The above image is the ``Table of Contents" Graphic: Graphical Image}

\begin{figure}
\centering
\includegraphics[width=3.5in, height=1.375in]{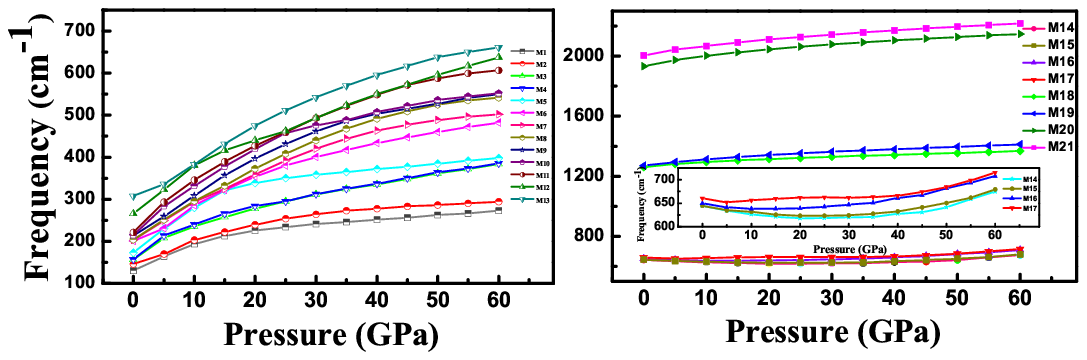}\\
\caption{(Colour online) Pressure-induced shifts of vibrational frequencies of LiN$_3$. The inset figure shows the enlarged case of pressure dependence of M14 to M17 modes.}
 \end{figure}


\begin{thebibliography}{}
\bibitem{Fair}
Fair, H. D.; Walker, R. F. Energetic Materilas; vol 1, Plenum Press; New York, 1977.

\bibitem{Bowden}
Bowden, F. P.; Yoffe, A. D. Fast Reactions in Solids; Butterworth Scientific Publications, London, UK, 1958.


\bibitem{Evans}
Evans, B.L.; Yoffe, A. D.; Gray, P. Chem. Rev. 1959, 59, 515.

\bibitem{Seel}
Seel, M.; Kunz, A.B. Int. J. Quantum. Chem. 1991, 39, 149.

 \bibitem{Gordienko}
Gordienko, A. B.; and Zhuravlev, Y. N.; and Poplavnoi, A. S. Phys. Status. Solidi B 1996, 198, 707.

 \bibitem{Gord}
Gordienko, A. B.; Poplavnoi, A. S. Phys. Status. Solidi B 1997, 202, 941.

 \bibitem{Younk}
Younk, E. H.; Kunz, A. B. Int. J. Quantum Chem. 1997, 63, 615.

\bibitem{Zhu}
Zhu, W.; Xiao, J.; Xiao, H. Chem. Phys. Lett. 2006, 422, 117.

\bibitem{Eremets}
Eremets, M. I.; Gavriliuk, A. G.; Trojan, I. A.; Dzivenko, D. A.; Boehler, R. Nat. Mater. 2004, 3, 558.

\bibitem{Erem}
Eremets, M. I.; Hemley, R. J.; Mao, H. K.; Gregoryanz, E. Nature 2001, 411, 170.

\bibitem{Ere}
Eremets, M. I.; Popov, M. Yu.; Trojan, I. A.; Denisov, V. N.; Boehler, R.; Hemley, R. J. J. Chem. Phys. 2004, 120, 10618.

\bibitem{Medvedev}
Medvedev, S. A.; Trojan, I. A.; Eremets, M. I.; Palasyuk, T.; Klap\"{o}tke, T. M.; Evers, J. J. Phys.: Condens. Matter, 2009, 21, 195404.

 \bibitem{Payne}
 Payne, M. C.; Teter, M. P.; Allan, D. C.; Arias, T. A.; Joannopoulos, J. D. Rev. Mod. Phys. 1992, 64, 1045.

 \bibitem{Segall}
 Segall, M.; Lindan, P.; Probert, M.; Pickard, C.; Hasnip. P.; Clark, S.; Payne, M. J. Phys.: Condens. Matter, 2002, 14, 271.

 \bibitem{Vanderbilt}
 Vanderbilt, D. Phys. Rev. B 1990, 41, 7892.

\bibitem{Kresse}
Kresse, G.; Furthmuller, J. Phys. Rev. B 1996, 54, 11169.

\bibitem{Fischer}
Fischer, T. H.; Almolf, J. J. Phys. Chem. 1992, 96, 9768.

\bibitem{Ceperley}
 Ceperley, D.M.; Alder, B. J. Phys. Rev. Lett. 1980, 45, 566.

\bibitem{PPerdew}
Perdew, J. P.; Zunger, A. Phys. Rev. B 1981, 23, 5048.

\bibitem{Perdew}
Perdew, J. P.; Burke, K.; Ernzerhof, M. Phys. Rev. Lett. 1996, 77, 3865.

\bibitem{Monkhorst}
Monkhorst, H. J.; Pack, J. Phys. Rev. B 1976, 13, 5188.

\bibitem{Pringle}
Pringle, G. E.; Noakers, D. E. Acta Crystallogr. B 1968, 24, 262.

\bibitem{Gonze}
Gonze, X. Phys. Rev. B 1997, 55, 10337.


\bibitem{Tulip}
 Refson, Keith.; Tulip, Paul R.; Clark, S. J. Phys. Rev. B 2006, 73,
155114.

\bibitem{Mehl}
 Mehl, M.J.; Osburn, J.E.; Papaconstantopoulus, D.A.; Klein, B.M. Phys. Rev. B 1990, 41, 10311.


\bibitem{Perger}
Perger, W. F. Int. J. Quant. Chem. 2010, 110, 1916.

\bibitem{Chevary}
 Perdew, John. P.; Chevary, J. A.; Vosko, S. H.; Jackson,Koblar A.;
      Pederson, Mark R.; Singh, D. J.; Fiolhais, C. Phys. Rev. B 1992, 46, 6671.

\bibitem{Mel}
Perdew, John P.; Levy, Mel. Phys. Rev. Lett. 1983, 51, 1884-1887.

\bibitem{Jones}
Jones, R. O.; Gunnarsson,  O.  Rev. Mod. Phys. 1989, 61, 689-746.

\bibitem{Errandonea}
Manjon, F.J.; Errandonea, D.; Segura, A.; Munoz, V.; Tobias, G.; Ordejon, P.; and Canadell, E. Phys. Rev. B 2001, 63, 125330.

\bibitem{Kanchana}

Kanchana, V.; Vaitheeswaran, G.; Souvatzis, P.; Eriksson, O.;  and Leb\`{e}gue, S. J.
Phys.: Condens. Matter, 2010, 22, 445402.

\bibitem{Bheem}
Bheema Lingam, Ch.; Ramesh Babu, K.; Tewari, Surya P.; Vaitheeswaran, G.; and Leb\`{e}gue, S. Phys. Status. Solidi RRL 5, No:1, 2010, 10.

\bibitem{Born}
Born, M.; Huang, K. Dynamical Theory of Crystal Lattices; Oxford University Press: Oxford, 1998.

\bibitem{Sinha}
Sinha, Sonali.; Sinha, T. P.; Mookerjee, A. Phys. Rev. B 2000, 62, 13, 8828.



%
%
%
%
%
%

%
%
%
%
%

%
%
%
%


%
%


%
%
%

%
%

\end{thebibliography}
\end{document}